\documentstyle[psfig]{mn}
\def\H0{{\it H}$_0$}

\def\q0{{\it q}$_0$}

\def\psqcm{cm$^{-2}$}

\title[BeppoSAX observations of GB1428+4217] 
{The blazar GB1428+4217: a warm absorber at z=4.72?}  
\author[A.C. Fabian, et al. ] 
{\parbox[]{6.5in} {A.C. Fabian$^1$, A. Celotti$^2$,
K. Iwasawa$^1$ and G. Ghisellini$^3$}\\ \\
$^1$Institute of Astronomy, Madingley Road, Cambridge CB3 0HA\\
$^2$SISSA, via Beirut, 2-4, 34014 Trieste, Italy\\ 
$^3$Osservatorio Astronomico di Brera-Merate, via Bianchi 46, 
23807 Merate (LC), Italy \\ }

\date{}

\begin{document}

\maketitle

\begin{abstract}
$Beppo$SAX observations of the high redshift ($z=4.72$) blazar
GB~1428+4217 confirm the presence of a complex soft X--ray spectrum
first seen with the ROSAT PSPC. Flattening below a rest frame energy
of 5 keV can be accounted for by absorption from an equivalent column
density of (cold) gas with $N_{\rm H} \sim 8\times 10^{22}$ cm$^{-2}$.
Below 2 keV a (variable) excess of a factor $\sim 20$ above the
extrapolated absorbed spectrum is also detected.  These findings are
consistent with and extend to higher redshifts the correlation between
increasing soft X--ray flattening and increasing $z$, previously
pointed out for large samples of radio--loud quasars.  We propose that
such features, including X--ray absorption, soft excess emission as
well as absorption in the optical spectra, can be satisfactorily
accounted for by the presence of a highly--ionized nuclear absorber
with column $N_{\rm H} \sim 10^{23}$ cm$^{-2}$, with properties
possibly related to the conditions in the nuclear regions of the host
galaxy. High energy X--ray emission consistent with the extrapolation
of the medium energy spectrum is detected up to $\sim 300$ keV (rest
frame).
\end{abstract}

\begin{keywords}galaxies: active - galaxies: individual: GB~1428+4217 
- X-ray: galaxies.
\end{keywords}

\section{Introduction}

High redshift Active Galactic Nuclei (AGN) are powerful tools to study
the physical and cosmological evolution of massive black holes and
their relationship with their galaxy hosts. A particularly interesting
class is formed by several high redshift ($z>4$), X-ray bright,
radio--loud quasars (Fabian et al. 1997, 1998; Moran \& Helfand 1997;
Zickgraf et al. 1997; Hook \& McMahon 1998) which present
characteristics typical of blazars. Interestingly, their X--ray
spectra have been recently found to systematically show the presence
of spectral flattening in the soft X--ray band. In GB~1428+4217
(Boller et al 2000), RXJ1028.6-0844 (Yuan et al 2000) and PMN
0525-3443 (Fabian et al 2000) such flattening implies, if interpreted
as due to intrinsic X-ray absorption by cold gas, column densities of
$1.5\times10^{22}$\psqcm, $2.1\times 10^{23}$\psqcm and $1.8\times
10^{23}$\psqcm, respectively.  These results appear to extend to
higher redshift the trend already found in nearer (up to $z=4.2$)
radio--loud quasars observed by ASCA and ROSAT (Cappi et al 1997,
Fiore et al. 1998, Reeves \& Turner 2000), the origin of which is
still unclear.

In particular for the most distant object, GB~1428+4217 at
$z=4.72$, previous observations in the X--ray and radio bands have
convincingly shown that this AGN is indeed a blazar (Fabian et
al. 1997, 1998).  Its spectral energy distribution (SED) appears
similar, although more extreme, to those of lower redshift blazars
of similar power and no strong evidence for different nuclear or jet
conditions has been found.

Here we present the results of $Beppo$SAX observations of this source.
They were in particular focused on: a) determining the amount and
shape of the flattening and thus its nature, as the MECS allows a good
determination of the medium energy X--ray spectrum; b) the detection
of the high energy PDS component, which constrains the position of
the $\gamma$--ray (inverse Compton) peak, which for GB~1428+4217 is
predicted -- by current models -- to be at or below $\sim$ MeV.

In the next section we present the analysis and results of the
$Beppo$SAX data, which are then discussed in Section 3.

\section{BeppoSAX data}

GB~1428+4217 was observed by $Beppo$SAX on 1999, Feb 4-7 (see Table
1). Signal has been detected in all three main instruments, the two
imaging instruments (LECS + MECS) and the PDS one (a collimator), with
exposure times of 27.3 (LECS), 90.3 (MECS) and 46.0 (PDS) ks.

A fit over the (1--10 keV) band with a simple power--law statistically
agrees with the previous ASCA results (Fabian et al. 1997), both in
spectral slope and in absolute flux. No evidence has been found of
spectral features associated with Fe emission around $\sim 1$ keV.
However, $Beppo$SAX data provide us with interesting information at
both the lower and higher energies.


\subsection{Low energy spectrum}

The extrapolation of such a power--law below 1 keV (and down to 0.4
keV) is well above the data, thus indicating either absorption in
excess of the Galactic one (column density $N_{\rm H}=1.4\times
10^{20}$ cm$^{-2}$) or an intrinsic flattening of the spectrum (see
Fig.~1). In Table~2 we report the results of the best fits with such
additional components: if intrinsic to the source the corresponding
column density is $\sim 7.8\times 10^{22}$ cm$^{-2}$. Errors are at
the 90 per cent confidence level for one parameter. The quality of the
data does not allow to statistically distinguish among these two
possibilities. The spectral slope and intensity are marginally
consistent (at 2$\sigma$) with those inferred from the ASCA and ROSAT
data (Fabian et al 1997, Boller et al 2000).

However, below 0.4 keV, there appears to be an excess above the model
(see Fig.~1). As this might be due to residuals in the background
subtraction, we performed a more careful analysis and in particular
re-extracted the spectrum by considering the local background. The low
energy excess is still present at the 2$\sigma$ level and we therefore
consider it, with reasonable probability, to be a real feature. This
emission has not been detected in the ROSAT data, taken $\sim$ 2
months before, suggesting that this component might be variable on
intrinsic timescales $\sim 10$ days. Note that (marginal) indications
of variability were also present during the ROSAT observation itself
(Boller et al 2000).

\begin{figure}
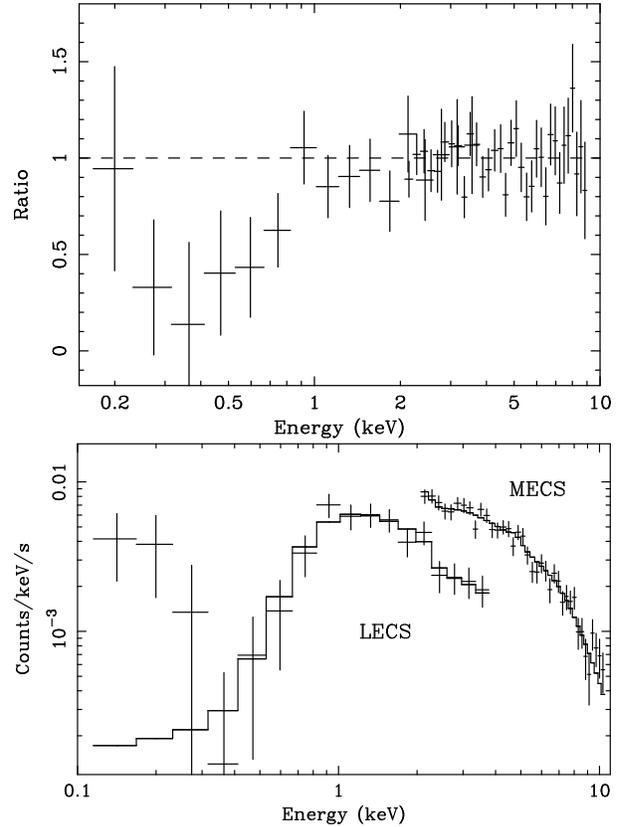

\centerline{\psfig{figure=1428_fig1a.ps,width=0.45\textwidth,angle=270}}
\centerline{\psfig{figure=1428_fig1b.ps,width=0.45\textwidth,angle=270}}
\caption{The ratio of the data to the absorbed power-law model, after
removal of the intrinsic but not the Galactic absorption components
(top panel); soft-to-medium X--ray BeppoSAX spectrum of GB1428+4217
fitted over the 1--10 keV band by a power-law with cold excess
absorption (bottom panel).}
\end{figure}

\begin{table}
\caption{\bf Log of the Observations}
\vskip 0.2 true cm
\begin{tabular}{l c c }
\hline \hline 

Object& Start time & End time \\

\hline

GB~1428+4217 & 99-Feb-04 18:04:47 & 99-Feb-07 01:26:40 \\

1ES1426+428 & 99-Feb-08 20:48:57 & 99-Feb-09 21:14:39 \\

\hline \hline
\end {tabular}
\end{table}

\begin{table*}
\caption{\bf Results of the spectral fits (0.4 --10 keV)}
\vskip 0.2 true cm
\begin{tabular}{l c c c c c c}
\hline \hline 

Model & $N_{\rm H}$ & $\Gamma_1$ & $E_{\rm break,obs}$ & $\Gamma_2$ &
$\chi^2$/d.o.f. \\

      & (10$^{22}$ cm$^{-2}$) & & (keV) & \\

\hline

Power-law + free intrinsic $N_{\rm H}$ (0.4-10 keV) & $7.8^{+8.7}_{-6.0}$ & 
$1.45\pm 0.10$ & -- & -- & 37.4/47 \\

Broken power--law (0.4-10 keV) & $N_{\rm H, gal}$ & $-1.5\pm 2.5$ & 
$0.86^{+1.68}_{-0.19}$ & $1.41\pm 0.09$ & 36.7/46 \\

\hline \hline
\end {tabular}
\end{table*}

\subsection{PDS data}

A strong signal has been also detected in the PDS band. However the
flux is well above the extrapolation of the LECS+MECS power--law (see
Fig.~2).  We thus checked for possible contamination in the PDS field
of view, and indeed found the presence of the BL Lac object
1ES1426+428 which, fortuitously, has been observed by SAX four days
after GB~1428+4217 (see Table~1). In collaboration with the
1ES1426+428 proposing team, we thus tried to disentangle the
contributions of the two sources in the PDS.

Also the spectrum of 1ES1426+428 is well described by a power-law
below 10 keV (Costamante et al 2000), with a photon index steeper than
that of GB1428+4217, $\Gamma\sim 1.93$. The two observed PDS spectra
are however very similar and significantly flatter than 1.9 (the data
from the 1ES1426+428 dataset are even slightly harder). Therefore this
implies that either the dominant contribution in the PDS comes from
GB1428+4217 (assuming its spectrum can be extrapolated from the
LECS+MECS power--law) or that at higher X--ray energies a flatter
component dominates in 1ES1426+428.

In the former case, a joint spectral fit to the datasets for
GB1428+4217 and 1ES1426+428 with simple power-law models requires
GB1428+4217 to vary a factor of $\sim 7$ between the two observations,
i.e. over a timescale as short as 2 d, while the high redshift quasar
did not show such strong and fast flux variations in the previous
observations (note that they correspond to an intrinsic timescale of
$\sim$ 8.4 hr).

In the second case, the hard PDS spectrum of the 1ES1426+428 dataset
is presumed to intrinsically flatten, similarly to that observed in
other BL Lac objects (e.g.  PKS2155--304, Chiappetti et al 1999), where
the (flat) inverse Compton component starts dominating over the
(steep) synchrotron one. We then estimated the contribution from
1ES1426+428 to the PDS data of the GB1428+4217 observation
iteratively, by assuming that the fluxes from the two sources are
constant between the observations and that the spectrum of GB1428+4217
is a single power-law across the $Beppo$SAX energy range (from 0.5 to
100 keV). Uncertainties in the relative normalization factors between
PDS and MECS are a problem: in order to reduce the number of free
parameters and, consistent with the assumption of constant fluxes,
we fixed these factors to 0.8 for both datasets.

Fig.~3 shows the 1ES1426+428 data with the best-fit power-law model for
the LECS+MECS extrapolated to the PDS energy range and with a
contribution from GB1428+4217 derived from its spectral fit (as in Fig.~2)
\footnote{We adopted a transmission efficiency of 46 per cent in the
PDS field of view during the 1ES1426+428 observation.}. Excess flux is
still present in the PDS band. The surplus emission is attributed to a
flat spectral component emerging above 10 keV from 1ES1426+428, i.e.
the entire broad band spectrum of the BL Lac is represented by a
broken power-law. Once the estimated contribution from GB1428+4217
($\Gamma= 1.45\pm 0.10$) is added this model gives a good fit to the
data ($\chi^2 = 242.2$ for 263 d.o.f., see Fig.~4).  Photon indices
below and above the break energy of 9.26 keV are $\Gamma_1\simeq 1.93$
and $\Gamma_2\simeq 1.34$, respectively.  The corresponding spectrum
for GB1428+4217 is well fitted with the single power-law ($\chi^2 =
40.8$ for 54 d.o.f., Fig.~5).  Although the contribution from
1ES1426+428 may be slightly overestimated (see the residuals in the
PDS band in Fig.~5), the PDS data in the GB1428+4217 observation
appear to be dominated by the contamination from 1ES1426+428.

The observer frame 2--10 keV fluxes for the two sources are
$2.76\times 10^{-12}$ erg cm$^{-2}$ s$^{-1}$ for GB1428+4217 and
$2.0\times 10^{-11}$ erg cm$^{-2}$ s$^{-1}$ for 1ES1426+428
(compatible with the results from the ASCA observation of 1994 Feb 6
by Kubo et al. 1998).  The 20--100 keV fluxes are $1.1\times 10^{-11}$
erg cm$^{-2}$ s$^{-1}$ and $6.3\times 10^{-11}$ erg cm$^{-2}$ s$^{-1}$
for GB1428+4217 and 1ES1426+428, respectively.

\begin{figure}
\centerline{\psfig{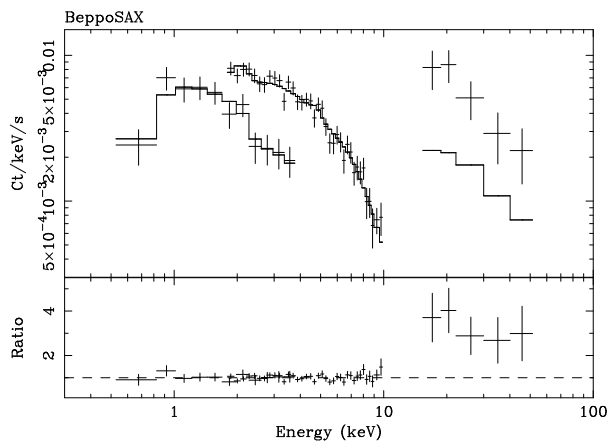}}
\caption{The BeppoSAX spectrum of GB~1428+4217. The plot shows the
best-fit power-law model to the LECS+MECS data. The surplus emission
in the PDS band is likely due to contamination from 1ES1426+428.}
\end{figure}

\begin{figure}
\centerline{\psfig{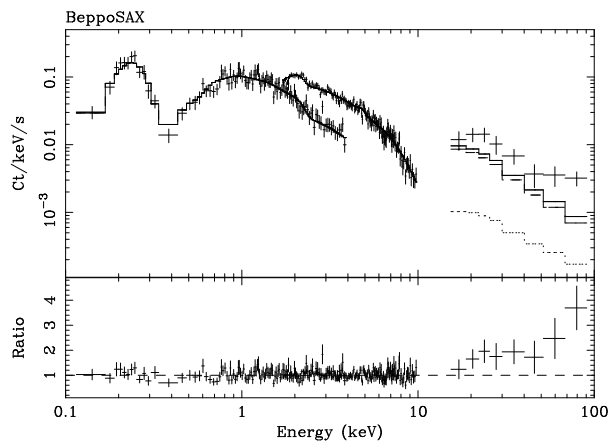}}
\caption{The spectrum of 1ES1426+428 with the best-fit power-law model
to the LECS+MECS data.  The contribution expected from GB~1428+4217 in
the PDS band has been taken into account (dotted line). There is still
surplus emission in the PDS band which is probably due to the BL Lac
object itself, unless GB~1428+4217 brightened by a factor of $\sim 7$
since it was observed with $Beppo$SAX two days before.}
\end{figure}

\begin{figure}
\centerline{\psfig{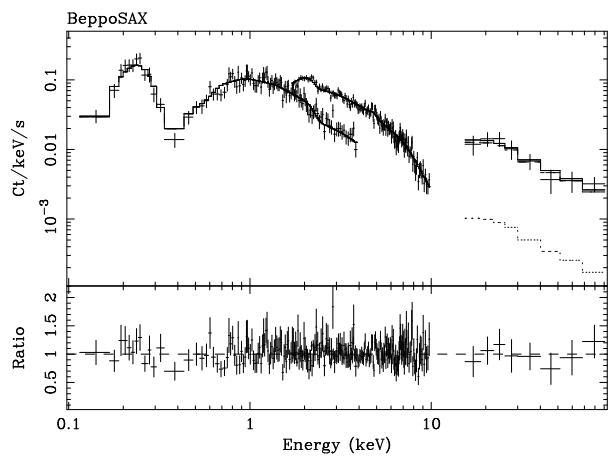}}
\caption{The spectrum of 1ES1426+428 fitted with a broken power-law
model. The plot shows the best-fit model including the contribution
from GB1428+4217 (see the caption of the previous figure).}
\end{figure}

\begin{figure}
\centerline{\psfig{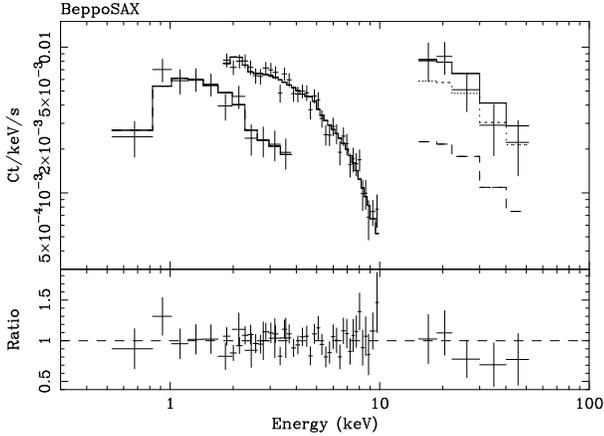}}
\caption{The spectrum of GB~1428+4217 fitted with a power-law model
including the contribution from 1ES1426+428 (dotted line) which is
based on the broken power-law fit (see previous figure). The
contribution from the BL Lac object appears to dominate the detected
PDS counts above 20 keV. }
\end{figure}

\section{Discussion}

We have presented the results of the $Beppo$SAX observations of the most
distant radio--loud X--ray luminous quasar GB~1428+4217. This has
allowed us to study and reveal interesting features at the extremes of
the $Beppo$SAX X--ray energy window.

At the higher energy end of the $Beppo$SAX broad band spectrum, the
contamination of the PDS data from 1E1426+428 unfortunately does not
allow us to determine the spectrum of GB~1428+4217 with sufficient
accuracy to constrain any curvature which might indicate the position
of the high energy $\gamma$--ray peak. However, from the
decontamination we performed, the spectrum appears consistently to
extend to the higher detected intrinsic energies, i.e. 280 keV in the
blazar rest frame.

\subsection{Low energy spectrum}

At the lowest energies, $< 1$ keV, the $Beppo$SAX observations
robustly confirmed the presence of a flattening in the spectrum found
in the ROSAT PSPC spectrum (Boller et al. 2000).  Secondly, below 0.4
keV emission a factor $\sim$ 20 in excess of the flattened spectrum
has been found (see Fig.~1). 

The origin of the flattening has been discussed in some detail by
several authors (Cappi et al 1997, Fiore et al 1998, Elvis et al 1998,
Boller et al 2000, Yuan et al 2000, Fabian et al 2000, Reeves \&
Turner 2000), but no definite conclusion could be drawn. Certainly,
the detection of such feature by different instruments (ROSAT, ASCA,
$Beppo$SAX) argues against any systematic mis-calibration effect.

The first interesting point which has been made is that the flattening
seems to be associated only with radio--loud objects, therefore
suggesting its origin to be intrinsic.  The number of good quality
X-ray spectra of radio--quiet quasars at $z> 1.5$ is unfortunately
limited.  However, the recent results by Vignali et al (2000),
combined with the findings by Reeves \& Turner (2000), imply that in
only 2 out of 15 radio--quiet quasars (in the range $z=1.8-2.5$) a
positive detection of flattening corresponding to $N_H \sim 10^{22}$
cm$^{-2}$ has been established, despite the biased selection of X--ray
loud sources.  For comparison, 5 out of 6 radio--loud objects in the
same redshift range have equivalent $N_H \ge 10^{22}$ cm$^{-2}$ (see
Fig.~6).  We therefore favour the possibility that the flattening is
due to an intrinsic property of the source, possibly associated with
the radio--loudness phenomenon. Nevertheless we note that radio-loud
objects tend to be more X--ray luminous and so yield the best spectra
at a given redshift and therefore in the following we also discuss the
possible role of absorption of intergalactic origin.

We consider it unlikely that large (galactic) scale gas is responsible
for the X--ray absorption, given the large masses of gas implied by
such hypothesis as well as the lack of any clear connection with the
radio--loud phenomenon.  Therefore in the following we concentrate on
the nuclear and/or cosmological properties which could account for
such feature.

A further key piece of information is the presence of a systematic
trend of the flattening in the spectra of radio--loud quasars to
increase with redshift (Cappi et al 1997, Fiore et al. 1998, Reeves \&
Turner 2000). The inclusion of the recent results on RXJ1028.6-0844
(Yuan et al 2000), PMN0525-3343 (Fabian et al 2000) and GB1428+4217
itself strengthens and extends this behavior up to $z\ge 4.7$, as
shown in Fig.~6.  Although no correlation between the flattening and
other spectral property has been previously found, we stress the
possible presence of a significant trend of increasing $N_{\rm H}$
with increasing hard X--ray (intrinsic 2--10 keV band) luminosity, as
can be argued from Fig.~7.  Unfortunately small statistics do
not allow us to disentangle the redshift and luminosity dependences.

In Fig.~6 we also show the line-of-sight value of $N_{\rm H}$ due to
the intergalactic medium (IGM) assuming solar abundances. If the IGM
was enriched by redshifts of about 4 to the same metallicity as
clusters of galaxies then the correlation with redshift could be
explained. As discussed in the ROSAT PSPC work on GB~1428+4217 (Boller
et al 2000) this conclusion does not agree with observations of the
metallicity of the Lyman $\alpha$ forest. Only if there was a strong
correlation between enrichment and temperature of the IGM phase might
some agreement occur, but even then the enrichment requirements would
be huge.

A more plausible interpretation of the apparent correlation shown in
Fig.~6 would then be that it arises from the detection limit for
absorption in present X-ray detectors. For example, a limit of a few
times $10^{20}$ cm$^{-2}$ at $z=0$ would corresponds to a few times
$(1+z)^3\, 10^{20}$ cm$^{-2}$ at redshift $z$, which roughly scales with
the points in Fig.~6.

\begin{figure}
\centerline{\psfig{figure=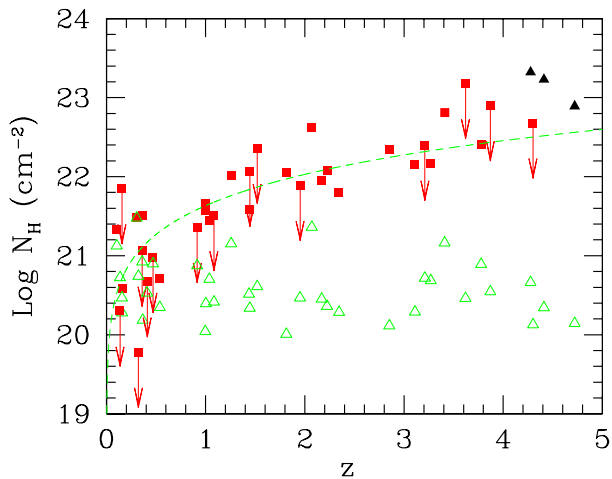,width=0.45\textwidth,angle=0}}
\caption{Absorption column density - as measure of the soft X--ray
flattening - vs redshift of radio--loud quasars. They comprise the
ASCA results reported by Reeves \& Turner (2000) (filled squares) and
the three sources above $z>4$ (GB~1428+4217, PMN0525-3343,
RXJ1028.6-0844, filled triangles). The empty symbols indicate the
Galactic column for each object. The dashed line corresponds to the
IGM column density, for $\Omega_0=0.3$, $\Omega_{\rm b}=0.06$ and
solar metallicity).}
\end{figure}
 
\begin{figure}
\centerline{\psfig{figure=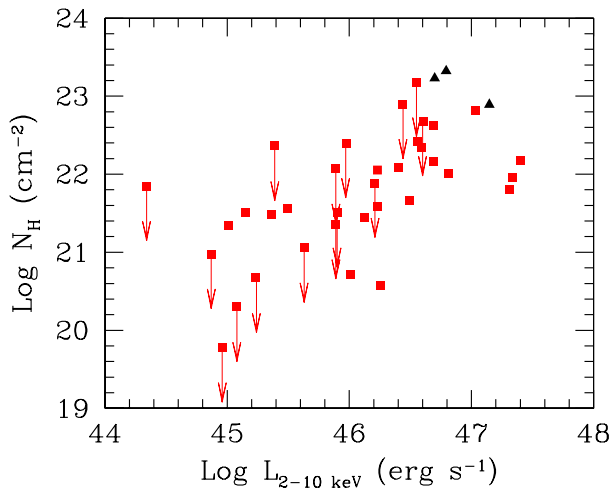,width=0.45\textwidth,angle=0}}
\caption{Absorption column density - as measure of the soft X--ray
flattening - vs hard X--ray luminosity (2--10 keV). Symbols and
objects as in Fig.~6.}
\end{figure}
 
\begin{figure}
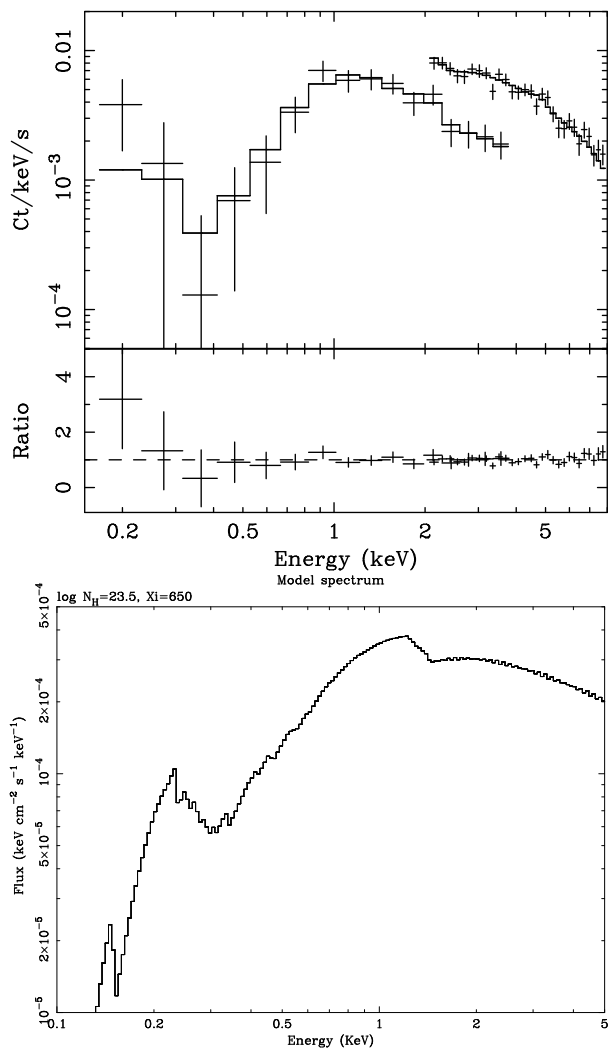

\centerline{\psfig{figure=1428_fig8a.ps,width=0.45\textwidth,angle=270}}
\centerline{\psfig{figure=1428_fig8b.ps,width=0.45\textwidth,angle=270}}
\caption{a) Fit with a warm absorber of the $Beppo$SAX data, for a
column density $N_{\rm H}=10^{23.5}$ cm$^{-2}$ and ionization
parameter $\xi =650$; b) the corresponding model. Edges at $\sim$ 1.5
keV (Fe), $\sim$ 0.25 (Mg) and 0.3 keV (Si) are clearly visible.
Solar metallicity has been assumed.}
\end{figure}
 
\begin{figure}
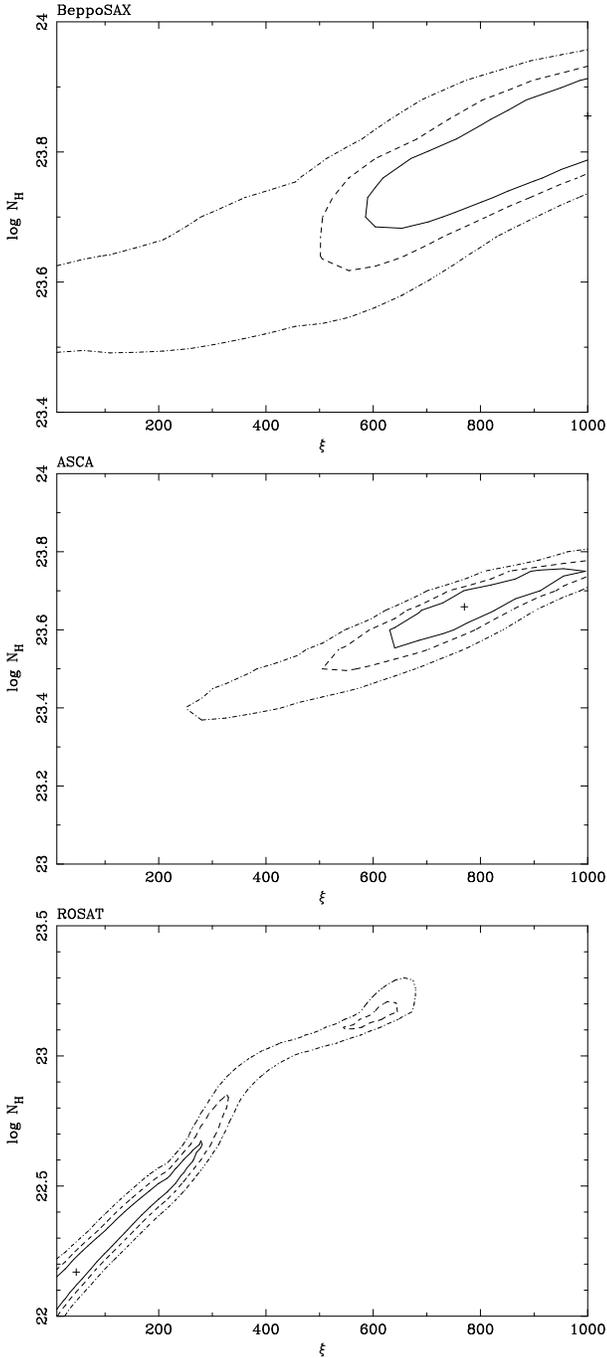

\centerline{\psfig{figure=1428_fig9a.ps,width=0.45\textwidth,angle=270}}
\centerline{\psfig{figure=1428_fig9b.ps,width=0.45\textwidth,angle=270}}
\centerline{\psfig{figure=1428_fig9c.ps,width=0.45\textwidth,angle=270}}
\caption{Confidence contours for the two parameters $N_{\rm H}$ and
$\xi$ describing the warm absorber properties, fitting $Beppo$SAX,
ASCA and ROSAT data of GB~1428+4217 (see references in the text). A
(marginally) consistent solution describing all the three datasets can
be found around $N_{\rm H}=10^{23.3-23.5}$ cm$^{-2}$ and $\xi \sim
600-700$.
}
\end{figure}

Let us now examine the alternative hypothesis of an intrinsic origin
of the flattening.  As discussed by Fabian et al (2000), on nuclear
scales this might arise because of: a) the characteristic shape of the
energy distribution of particles and/or photons which are believed to
be responsible -- through inverse Compton scattering on soft external
photon field -- for the high energy spectral component in blazars; b)
absorption by nuclear, highly ionized, gas.

Let us therefore consider in turn how these two possibilities can
account for the spectral characteristics discussed so far. These,
including the results of the present analysis on GB~1428+4217, can be
summarized as: a spectral flattening below $\sim$ 5 keV, soft excess
emission below $\sim$ 2.3 keV, a possible connection with
radio--loudness/jets and with the redshift and/or X--ray luminosity,
an optical spectrum showing evidence for absorption features (Ly
$\alpha$-NV and CIV) (Hook \& McMahon 1998; see Fabian et al 2000 for
the similar case of PMN 0525-3433).  We recall that the latter
(optical) feature is indeed typically seen in outflows (Brandt, Laor
\& Wills 2000).

Within scenario a), which implicitly links the observed properties
with blazar--type sources, the flattening could be caused by several
effects, such as: an intrinsic break in the low energy part of the
non--thermal electron distribution (e.g. caused by incomplete cooling)
which would require a break energy $\sim$ few $m_{\rm e}$ c$^2$; a
paucity in the (relative) amount of synchrotron photons which are
scattered at higher energies (synchrotron self--Compton emission) and
thus typically contribute to the soft-to-medium X--ray band flux,
filling the `depression' between the synchrotron and inverse Compton
components in blazars (e.g. Sikora, Begelman \& Rees 1994)
\footnote{Note that an increase in the dominance of the external
Compton component would be also naturally associated with an increase
in the source X--ray luminosity.}; the intrinsic flattening of the
external soft photon energy distribution below its peak, which
translates into a flattening in the scattered inverse Compton
component at soft X--ray energies.  Also the presence of a soft X--ray
excess, as seen in GB~1428+4217 and RXJ1028.6-0844 could be consistent
with this scenario as due to either Comptonization of a cold electron
component in the jet (bulk Compton emission, Sikora et al. 1994) or
the high energy tail of the soft radiation field.

Nevertheless a serious problem arises with this scenario, if the soft
X--ray spectrum is produced by low energy electrons. In fact, while
the uncertainties in the determination of the spectral
shape/absorption in the high redshift objects does not allow the
presence of any spectral variation to be found, soft X--ray flux
variability has been established for GB1428+4217 on timescales $\sim$
1 d (Fabian et al. 1999).  Although strong and fast variability is of
course expected in blazars thanks to the enhancement caused by
relativistic beaming, the cooling timescales of electrons emitting at
soft X--ray energies - as discussed in scenario a) - would greatly
exceed the observed variability timescale.

In scenario b), i.e.  absorption by small scale gas, such as nuclear
absorbing winds/outflows, the flattening {\it and} the soft X--ray
excess can be successfully accounted for by the presence of a
warm/highly ionized absorber with a column density of the order of
$10^{23}$ cm$^{-2}$ and $\xi\sim 10^2$ (where the ionization parameter
is defined as $\xi \equiv L /(n R^2)$; $L$ is the ionizing luminosity
from 1 to 1000~Ryd).  In Fig.~8 we report the results of the spectral
fitting to the $Beppo$SAX data with such a model, as calculated using
the photoionization code CLOUDY C90.04 (Ferland et al. 1998). The
predicted model spectrum (see Fig.~8) clearly shows the presence of
edges at $\sim$ 1.5 keV (FeXX-FeXXIV) and $\sim0.25-0.3$ keV (Mg).
Most absorption is due to oxygen. Solar metallicity has been assumed
in these nuclear regions. $\chi^2=29.2/40$.

We further explored the viability of such scenario by considering
previous observations of GB~1428+4217.  In Fig.~9 the contour plots
for the two critical parameters of the model -- the gas column density
and the ionization state -- are shown for the BeppoSAX, ASCA and ROSAT
PSPC data (Fabian et al. 1997, 1998).  A common range of (marginally
consistent) parameters, around $\xi \sim 600-700$ and column densities
$\sim 3\times 10^{23}$, can be found for the three datasets. Some of
the variation may be due to spectral variability among the
observations.

Note that if this picture is correct the lack of significant
extinction in the optical--UV bands (similar to what seen in PMN
0525-3433) also requires that the absorbing gas is highly ionized
and/or rich in metals and lacking of dust.  (The observed intense and
fast variability is inconsistent with being caused by the postulated
warm absorber itself and therefore has to be ascribed to the primary
emission from relativistic plasma.)

The ionization parameter $\xi$, column density $N_{\rm H}$ and
luminosity $L$ along our line of sight to GB~1428+4217 require the
warm absorber to be located at an estimated distance from the nucleus
of less than about 200~pc.  The increase of the amount of absorption
with redshift but also with luminosity, argues for an even larger
increase in the amount of gas, plausibly associated with a large
(super--Eddington) accretion rate of infalling gas in the early stages
of nuclear activity, and/or outflows possibly confining powerful radio
jets. The warm absorber may also be the inner part of the dense
interstellar medium of the young host galaxy (see Fabian 1999 for a
more general discussion).

If this gas completely envelopes the source then its mass is $\sim
10^{9} r_{200}^2$ $M_{\odot}$, where the radius is $200r_{200}$ pc. It
is possible that in directions well away from the axis of the blazar
jet the surrounding gas is of low ionization and so opaque to all but
hard X-rays and that only in the jet direction the X-ray intensity is
powerful enough to ionize the gas.  Therefore in more extreme/high $z$
sources, such gas might completely hide the presence of activity in
the soft-to-medium X--ray band.  Under such conditions of high optical
depths also the nuclear disk radiation might be partly reflected back
by the gas, thus enhancing the nuclear radiation energy density.  The
high amount of scattering gas should also allow significant detection
of nuclear emission in the most powerful radio galaxies.

The decrease in absorption seen in objects below a redshift $\sim 1$
(see Fig.~6) may be associated with this gas being consumed by star
formation, or having been blown out of the host by a quasar wind. The
predominance of absorption in radio--loud objects could be related to
them having more massive host galaxies (and thus it being more
difficult to blow the gas away) and/or to the radio phase being
associated with the quasar youth.

Observations of GB~1428+4217 and similar high redshift blazars with
high sensitivity and energy resolution instruments, such as those
provided by XMM, will soon allow us to determine whether these conditions
are ubiquitous in high $z$ sources. They should robustly test the
proposed warm absorber scenario by detecting the edges predicted in
Fig.~8 and so enable us to infer the physical conditions in the active
nuclear regions of primeval galaxies.

\noindent
\section{\bf Acknowledgments}

We thank Gary Ferland for the use of his code CLOUDY.  The Royal
Society (ACF) and the Italian MURST and ASI grant ARS-99-74 (AC) are
thanked for financial support.

\end{document}